\documentclass[aps,prb,twocolumn,showpacs,amsmath,floatfix,citeautoscript]{revtex4}
\usepackage{graphicx}
\usepackage{graphics}
\usepackage{dcolumn}
\usepackage{hyperref}
\hypersetup{colorlinks,citecolor=blue,linkcolor=blue,urlcolor=blue}

\begin{document}
\title{A Systematic Study of Chloride Ion Solvation in Water using van der Waals Inclusive Hybrid Density Functional Theory}
\author{Arindam Bankura$^{1}$}
\author{Biswajit Santra$^{2}$, Robert A. DiStasio Jr.$^{2}$}
\author{Charles W. Swartz$^{1}$, Michael L. Klein$^{1}$}
\author{Xifan Wu$^{3}$}
\affiliation{$^{1}$ Institute for Computational Molecular Science and Department of Chemistry, Temple University, Philadelphia, PA 19122, USA}
\affiliation{$^{2}$ Department of Chemistry, Princeton University, Princeton, NJ 08544, USA} 
\affiliation{$^{3}$ Department of Physics, Temple University, Philadelphia, PA 19122, USA}

\date{\today}

\begin{abstract}
In this work, the solvation and electronic structure of the aqueous chloride ion solution was investigated using Density Functional Theory (DFT) based \textit{ab initio} molecular dynamics (AIMD). From an analysis of radial distribution functions, coordination numbers, and solvation structures, we found that exact exchange ($E_{\rm xx}$) and non-local van der Waals (vdW) interactions effectively \textit{weaken} the interactions between the Cl$^-$ ion and the first solvation shell. With a Cl-O coordination number in excellent agreement with experiment, we found that most configurations generated with vdW-inclusive hybrid DFT exhibit 6-fold coordinated distorted trigonal prism structures, which is indicative of a significantly disordered first solvation shell. By performing a series of band structure calculations on configurations generated from AIMD simulations with varying DFT potentials, we found that the solvated ion orbital energy levels (unlike the band structure of liquid water) strongly depend on the underlying molecular structures. In addition, these orbital energy levels were also significantly affected by the DFT functional employed for the electronic structure; as the fraction of $E_{\rm xx}$ was increased, the gap between the highest occupied molecular orbital of Cl$^-$ and the valence band maximum of liquid water steadily increased towards the experimental value.
\end{abstract}

\maketitle

\section{Introduction}

The nature of the interaction between the hydrogen bond (HB) network of liquid water and the solvated chloride ion (a member of the Hofmeister series) is currently a topic under intense research due to its fundamental importance in biochemistry, atmospheric chemistry, and geological processes~\cite{Abbatt_2006,Knipping_2000,Spicer_1998,Beekman_2011}.
Experimentally, the interaction between the solvated chloride ion, Cl$^-$, and the surrounding aqueous environment has been successfully probed by a variety of techniques~\cite{Cummings_1980,Delahay_1982,Ghosal_2005,Copestake_1985,Yamagami_1995,Kropman_2001,Kropman_2002,Bakker_2005,Megyes_2006,Megyes_2008,Tongraar_2010,
Dang_2006,Bruni_2012,Mancinelli_2007_2,Botti_2004,Soper_2006,Skinner_2012,Mancinelli_2007,Winter_2005,Winter_2006,Seidel_2011}. 
In particular, X-ray and neutron scattering as well as X-ray absorption measurements, have been used to investigate the solvation structure of aqueous chloride solutions, providing experimental determination of the Cl-O and Cl-H radial distribution functions (RDF), $g_{\rm Cl-O}(r)$ and $g_{\rm Cl-H}(r)$, and the associated coordination number, \textit{i.e.}, the number of water molecules populating the first solvation shell surrounding a given Cl$^-$ ion~\cite{Cummings_1980,Copestake_1985,Yamagami_1995,Megyes_2006,Megyes_2008,Tongraar_2010,Dang_2006,Bruni_2012,Mancinelli_2007,Mancinelli_2007_2,Soper_2006}.
In addition, the electronic structure associated with this fundamental aqueous ionic solution was studied by Delahay~\cite{Delahay_1982} and Winter {\it et al.}~\cite{Winter_2006}, in which state-of-the-art photoemission spectroscopy (PES) was utilized to characterize the valence 3$p$ bands of the solvated Cl$^-$ ion, which were found to be approximately 1.25--1.50 eV above the valence band maximum (VBM) of liquid water.

From a theoretical and computational point of view, first-principles based computational methods such as \textit{ab initio} molecular dynamics (AIMD) have become powerful tools in the study of condensed-phase systems such as liquids~\cite{laasonen_jcp_1993,tuckerman_jcp_1995} and aqueous ionic solutions~\cite{Ikeda_2003,Heuft_2003,Tongraar_2003,Mallik_2008,Scipioni_2009,Guardia_2009,Calemana_2011,Leung_2009,Bankura_2013,Ge_2013,Zhang_2013,Gaiduk_2014}.
With the AIMD technique, the nuclear potential energy surface is generated ``on the fly'' from the electronic ground state~\cite{Car_1985} without the need for empirical input, thereby allowing for a quantum mechanical treatment of not only the structure and dynamics of a given molecular system of interest, but also its electronic and dielectric properties, as well as potential chemical reactions~\cite{Zipoli_2010,geissler_science_2001,ikeshoji_pccp_2009} (\textit{i.e.}, the breaking and forming of chemical bonds).
Since the initial pioneering simulations of liquid water,~\cite{laasonen_jcp_1993,tuckerman_jcp_1995} AIMD has been applied to many complex problems in biology, chemistry, and energy research, \textit{e.g.}, the designing of efficient catalysts for hydrogen production~\cite{Zipoli_2010} and the modeling of the auto-ionization~\cite{geissler_science_2001} and electrocatalytic splitting of water,~\cite{ikeshoji_pccp_2009} to name a few.

In particular, AIMD simulations employing density functional theory (DFT) as the source of the underlying quantum mechanical potential have been used to generate reasonably accurate microscopic descriptions of the structure of the aqueous chloride ion solution,~\cite{ Ikeda_2003,Heuft_2003,Tongraar_2003,Mallik_2008,Scipioni_2009,Leung_2009,Guardia_2009,Calemana_2011,Bankura_2013,Ge_2013,Zhang_2013} with RDFs and associated coordination numbers that were consistent with the available experimental data.
These studies found that the predicted solvation structure in the aqueous chloride ion solution is moderately dependent on the exchange-correlation (XC) potential, with semi-local generalized-gradient approximation~\cite{Perdew_1992,Becke_1992} (GGA) functionals such as PBE~\cite{Perdew_1996} and BLYP~\cite{Becke_1988,Lee_1988} yielding similar predictions to hybrid functionals such as PBE0.~\cite{Barone_1999}
In addition, the HB network of water was found to be only locally perturbed by the presence of the solvated ion, a finding which is consistent with the general expectation that Cl$^-$ is a weak disrupter of the aqueous environment~\cite{Mancinelli_2007,Mancinelli_2007_2,Soper_2006}.
However, these studies also demonstrated that the electronic structure of this aqueous ionic solution is very sensitive to the adopted XC approximation~\cite{Ge_2013,Zhang_2013}.
In particular, higher-level electronic structure calculations performed on GGA-DFT generated configurations (structures) were found to significantly underestimate the energy levels of the solvated Cl$^-$ ion, with resulting energetics that can even be qualitatively incorrect~\cite{Ge_2013,Zhang_2013}.

In the study of an aqueous ionic solution, an accurate theoretical description of the underlying HB network is a prerequisite for characterizing and understanding the interaction between a solvated or embedded ion and its surrounding environment.
In neat liquid water, the predictive power of DFT-based AIMD in the microscopic description of the HB network critically depends on the accuracy of the underlying XC functional utilized in the quantum mechanical treatment of the electronic degrees of freedom.
In this regard, it is now clear that the most widely used XC functionals, \textit{i.e.}, the class of functionals based on the GGA, have severe limitations when applied to liquid water~\cite{asthagiri_pre_2003,grossman_jcp_2004,schwegler_jcp_2004,fernandez_jcp_2004,kuo_jpcb_2004,mcgrath_cpc_2005,vandevondele_jcp_2005,sit_jcp_2005,fernandez_ms_2005,mcgrath_mp_2006,lee_jcp_2006,Todorova_2006,lee_jcp_2007,guidon_jcp_2008,kuhne_jctc_2009,mattson_jctc_2009,yoo_jcp_2009,Zhang_2011,bartok_prb_2013,alfe_jcp_2013,lin_jpcb_2009,jonchiere_jcp_2011,wang_jcp_2011,zhang_jctc_2011,mogelhoj_jpcb_2011,lin_jctc_2012,yoo_jcp_2012,schmidt_jpcb_2009,ma_jcp_2012,DelBen_2013} as well as the crystalline phases of ice.~\cite{feibelman_pccp_2008,santra_prl_2011,santra_jcp_2013,labat_jcc_2011,kambara_pccp_2012,murray_prl_2012,fang_prb_2013,macher_jcp_2014}
Most notably, GGA-DFT suffers from the presence of self-interaction error (SIE)~\cite{Shore_1977,perdew_prb_1981} and the neglect of non-local electron correlation effects that are responsible for van der Waals (vdW) or dispersion interactions; as a result, these deficiencies in the GGA XC potential manifest as significant overstructuring and excessively sluggish dynamics in aqueous systems such as ambient liquid water.
Beyond the choice of the XC functional, most AIMD simulations of liquid water and aqueous ionic solutions performed to date have adopted classical mechanics for the nuclear equations of motion and therefore completely neglect nuclear quantum effects (NQE)---another approximation that has been deemed insufficient for a highly accurate quantitative description of the microscopic structure and HB network in aqueous systems.
In the case of liquid water, light atoms such as hydrogen deviate significantly from classical behavior even at room temperature,~\cite{chen_prl_2003,morrone_prl_2008,ceriotti_pnas_2013} as evidenced by experimental isotope effect studies which demonstrated a softening of the liquid structure (\textit{i.e.}, in the comparison of H$_2$O to D$_2$O).~\cite{soper_prl_2008} 
Hence the neglect of NQE in aqueous systems such as ambient liquid water leads to overstructuring in the predicted RDFs.~\cite{morrone_prl_2008,ceriotti_prl_2012,paesani_jcp_2007,fanourgakis_jcp_2006}

A commonly adopted method to alleviate the deleterious effects of SIE in GGA-DFT is the use of hybrid XC functionals, wherein a fraction of exact (or Hartree-Fock) exchange ($E_{\rm xx}$) is included in the density functional approximation.
Due to the relatively high computational cost associated with these XC functionals, applications of hybrid DFT have mostly been restricted to small gas-phase clusters of water,~\cite{santra_jcp_2007,santra_jcp_2008,santra_jcp_2009,wang_jcp_2010,gillan_jcp_2012} although recently hybrid functionals have been applied in the study of several crystalline phases of ice~\cite{santra_prl_2011,santra_jcp_2013,labat_jcc_2011,kambara_pccp_2012,erba_jpcb_2009} and liquid water.~\cite{zhang_jctc_2011,Zhang_2011,Todorova_2006,guidon_jcp_2008,DelBen_2013,Distasio2014}
In comparison to GGAs, these studies demonstrated that the energetic, structural, and vibrational properties of these aqueous systems, as predicted by hybrid DFT calculations, are generally in closer agreement with the available experimental data.~\cite{santra_jcp_2009,wang_jcp_2010,gillan_jcp_2012,santra_prl_2011,labat_jcc_2011,kambara_pccp_2012,erba_jpcb_2009,santra_jcp_2013,Distasio2014}
Indeed, an accurate microscopic description of the HB network by hybrid XC functionals has a non-negligible effect on the theoretical characterization of the solvated chloride ion. In this regard, it was found that the use of hybrid DFT was crucial in obtaining a qualitatively correct energy difference between Cl$^-$ and the VBM of liquid water.
In addition, both hybrid and GGA XC functionals lack the ability to describe vdW/dispersion interactions, which arise from non-local dynamical electron correlation and have a substantial effect on the microscopic structure of condensed-phase aqueous systems.
In fact, the explicit inclusion of vdW interactions in DFT has been shown to significantly improve upon the theoretical description of the transition pressures among the high-pressure phases of ice~\cite{santra_prl_2011,kambara_pccp_2012,murray_prl_2012} and the predicted equilibrium density of liquid water.~\cite{schmidt_jpcb_2009,ma_jcp_2012,DelBen_2013}
While many recent studies have concluded that the structure of liquid water significantly softens when vdW interactions are accounted for in the underlying XC potential, the extent to which these non-local forces affect the structure of liquid water is largely dependent upon the given approach utilized to facilitate vdW-inclusive DFT.~\cite{lin_jpcb_2009,jonchiere_jcp_2011,wang_jcp_2011,zhang_jctc_2011,mogelhoj_jpcb_2011,lin_jctc_2012,yoo_jcp_2012,schmidt_jpcb_2009,ma_jcp_2012,DelBen_2013,Distasio2014}
Hence, the inclusion of vdW or dispersion interactions in the underlying XC potential is expected to have a non-negligible effect on the theoretical description of the HB network and must be accounted for in the study of aqueous ionic solutions such as the solvated chloride ion.

Recently it was shown that utilization of the hybrid PBE0 functional,~\cite{perdew_jcp_1996,Barone_1999} which includes 25\% exact exchange, in conjunction with a fully self-consistent (SC) implementation of the density-dependent vdW/dispersion correction of Tkatchenko and Scheffler~\cite{Tkatchenko_2009} (TS-vdW),~\textit{i.e.}, the PBE0+TS-vdW(SC) XC functional, yields an oxygen-oxygen structure factor, $S_{\rm OO}(Q)$, and corresponding RDF, $g_{\rm OO}(r)$, that are in quantitative agreement with the best available experimental data.~\cite{Distasio2014,Skinner2013}
This level of agreement between~\textit{ab initio} simulations and experiment was attributed to an increase in the relative population of water molecules in the interstitial region (\textit{i.e.}, the region between the first and second coordination shells), a collective reorganization in the liquid phase which is facilitated by a weakening of the HB strength by the use of a hybrid XC functional, coupled with a relative stabilization of the resultant disordered liquid water configurations by the inclusion of long-range vdW/dispersion interactions.
In fact, this increasingly more accurate description of the underlying HB network in liquid water also yielded other correlation functions, such as the oxygen-hydrogen RDF, $g_{\rm OH}(r)$, and the higher-order oxygen-oxygen-oxygen triplet angular distribution, $P_{\rm OOO}(\theta)$, which encodes the degree of local tetrahedrality, as well as electrostatic properties, such as the effective molecular dipole moment, that are in much better agreement with experiment.~\cite{Distasio2014}
In this regard, the overall agreement between experiment and the PBE0+TS-vdW(SC) description of the microscopic structure of ambient liquid water is indeed a very promising starting point for accurately investigating the structural and energetic properties that characterize the aqueous chloride ion solution, which is the main focus of the work reported herein.

\section{Computational Details}

In this work, we have systematically performed a series of Car-Parrinello AIMD simulations~\cite{Car_1985} of the aqueous chloride ion solution at ambient conditions using a hierarchy of different XC functionals.
The sequence of XC functionals employed herein includes the standard semi-local GGA of Perdew, Burke, and Ernzerhof (PBE),~\cite{Perdew_1996} the corresponding hybrid PBE0~\cite{perdew_jcp_1996,Barone_1999} which includes 25\% exact exchange, and the self-consistent (SC) dispersion-corrected analogs~\cite{distasio_unpublished} thereof, \textit{i.e.}, PBE+TS-vdW(SC) and PBE0+TS-vdW(SC), based on the Tkatchenko-Scheffler~\cite{Tkatchenko_2009} density-dependent vdW/dispersion functional.

All of these AIMD simulations were performed in the canonical ($NVT$) ensemble using periodic simple cubic simulation cells containing Cl$^-$ ion in (H$_2$O)$_{63}$ with lattice parameters set to reproduce the experimental density of liquid water at ambient conditions.
All of the AIMD simulations were initially equilibrated for approximately 6 ps and then continued for an additional 20--50 ps for data collection (with the AIMD simulations employing the hybrid PBE0 functional at the lower end of this range of simulation times).
Since a classical treatment of the nuclear degrees of freedom was found to be insufficient for a quantitatively accurate description of the microscopic structure of ambient liquid water, we have performed all AIMD simulations in this work at the elevated temperature of 330 K, a technique that is suggested by the lowest-order perturbative expansion of the free energy (in $\hbar$) to account for the quantum mechanical nature of the nuclear degrees of freedom.~\cite{landau_sm_book_1969}
In practice, this increase of approximately 30 K in the simulation temperature has been found to mimic the nuclear quantum effects (NQE) in structural quantities such as the oxygen-oxygen radial distribution function ($g_{\rm OO}(r)$) in both DFT~\cite{morrone_prl_2008} and force field~\cite{paesani_jcp_2007,fanourgakis_jcp_2006} based MD simulations of liquid water.

All calculations reported herein were performed within the plane-wave and pseudopotential framework and utilized a modified development version of the Quantum ESPRESSO (QE) software package.~\cite{QE-2009}
To meet the additional computational demands associated with large-scale AIMD simulations based on hybrid XC functionals, we have employed a linear scaling O($N$) exact exchange algorithm that exploits the natural sparsity associated with the real-space maximally localized Wannier function (MLWF)~\cite{marzari_prb_1997} representation of the occupied Kohn-Sham electronic states.~\cite{wu_prb_2009,ko_unpublished}
In addition, we have also developed and utilized a linear scaling O($N$) self-consistent implementation of the TS-vdW dispersion correction,~\cite{distasio_unpublished} which provides a framework for computing atomic $C_6$ dispersion coefficients as explicit functionals of the charge density, \textit{i.e.}, $C_{6,AB}=C_{6,AB}[\rho(\mathbf{r})]$, thereby accounting for the local chemical environment surrounding each atom.~\cite{Tkatchenko_2009}
More explicit descriptions of the theoretical methods employed herein can be found in Ref. [~\onlinecite{Distasio2014}].

The Car-Parrinello (CP)~\cite{Car_1985} equations of motion for the nuclear and electronic degrees of freedom were integrated using the standard Verlet algorithm and a time step of 4.0 a.u. ($\approx$ 0.1 fs).
To ensure an adiabatic separation between the electronic and nuclear degrees of freedom in the CP dynamics, we used a fictitious electronic mass of 300 a.u., which was found to be a reasonable choice for the simulation of water,~\cite{grossman_jcp_2004} and the nuclear mass of deuterium for each hydrogen atom.
All electronic wavefunctions were expanded using a plane wave basis set with a kinetic energy cutoff of 72 Ry, and the interactions between the valence electrons and the ions (consisting of the nuclei and their corresponding frozen-core electrons) were treated with Troullier-Martins type norm-conserving pseudopotentials.~\cite{troullier_prb_1991}
Ionic temperatures were controlled with massive Nos\'{e}-Hoover chain thermostats,~\cite{Nose_1984,Hoover_1985} each with a chain length of 4.~\cite{martyna_jcp_1992}
A neutralizing background charge was included in the Ewald summation of the electrostatic energy to compensate for the net negative charge of the chloride ion solution.

To examine the electronic structure surrounding the solvated chloride ion, we have performed wavefunction optimizations (\textit{i.e.}, self-consistent solutions of the non-linear Kohn-Sham equations) using the PBE, PBE0 (25\% exact exchange), and BHLYP~\cite{Becke_1993} (50\% exact exchange) XC functionals on a set of configurations (structures) selected at even intervals along a given MD trajectory.
Real-space integration of the orbital densities surrounding the solvated chloride ion was performed within a sphere of radius $R=1.5$ \AA\ to determine the ionic contribution to the overall electronic structure and to identify the highest-occupied molecular orbital (HOMO) states.
All PBE and PBE0 wavefunction optimizations were performed with the QE software package.~\cite{QE-2009}
Calculations with the BHLYP XC functional were performed with the CP2K suite of programs~\cite{cp2k} using a split-valence triple-$\zeta$ basis set (appended with an additional set of polarization functions) in conjunction with Goedecker-Teter-Hutter type pseudopotentials~\cite{Goedecker_1996} and a density cut-off  of 400Ry.

\section{Results and Discussions}

\subsection{Solvation Structure of the Aqueous Cl$^-$ Ion \label{sec:solvation}}

\begin{figure}[t!]
\vspace{0.5 cm}
\begin{center}
\includegraphics[width=8.0cm]{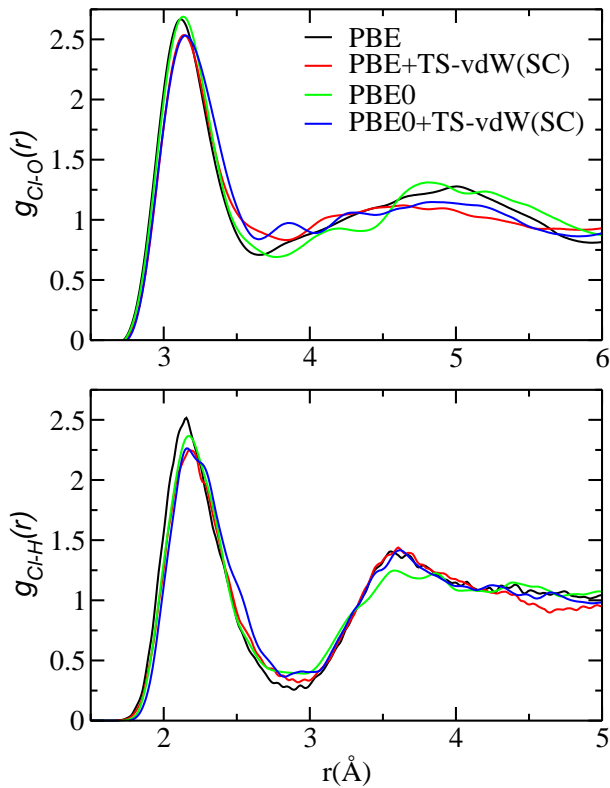}
\caption{The chlorine--oxygen (upper panel) and chlorine--hydrogen (lower panel) radial distribution functions, $g_{\rm Cl-H}(r)$ and $g_{\rm Cl-O}(r)$, of the aqueous chloride ion solution obtained from theory \textit{via} the DFT-based AIMD simulations performed in this work.}
\label{rdf1} 
\end{center}
\end{figure}

\begin{table*}[ht]
\vspace{-0.5 cm}
\caption{Numerical values for the intensities ($h_{\rm Cl-O}^{\rm max}$ and $h_{\rm Cl-H}^{\rm max}$) and positions ($r_{\rm Cl-O}^{\rm max}$ and $r_{\rm Cl-H}^{\rm max}$ in \AA) of the first maximum in the RDFs depicted in Fig.~\ref{rdf1}, the positions ($r_{\rm Cl-O}^{\rm min}$ and $r_{\rm Cl-H}^{\rm min}$ in \AA) of the first minimum in the RDFs depicted in Fig.~\ref{rdf1}, as well as the corresponding coordination numbers ($n_{\rm Cl-O}$ and $n_{\rm Cl-H}$). Available experimental findings for these quantities have been compiled in the bottom rows.}
\label{table1}
\begin{center}
\begin{tabular}{c|cccc|cccc}
\hline\hline
  Method & $h_{\rm Cl-O}^{\rm max}$ & $r_{\rm Cl-O}^{\rm max}$ & $r_{\rm Cl-O}^{\rm min}$ & $n_{\rm Cl-O}$ & $h_{\rm Cl-H}^{\rm max}$ & $r_{\rm Cl-H}^{\rm max}$ & $r_{\rm Cl-H}^{\rm min}$ & $n_{\rm Cl-H}$ \\
\hline
PBE             & 2.67$\pm$0.11 & 3.11$\pm$0.02 & 3.64$\pm$0.03 & 5.5$\pm$0.2 & 2.51$\pm$0.16 & 2.14$\pm$0.03 & 2.88$\pm$0.06 & 5.2$\pm$0.1   \\
PBE+TS-vdW(SC)  & 2.53$\pm$0.13 & 3.14$\pm$0.01 & 3.78$\pm$0.12 & 6.3$\pm$0.9 & 2.25$\pm$0.15 & 2.17$\pm$0.02 & 2.93$\pm$0.06 & 5.2$\pm$0.3   \\
PBE0            & 2.70$\pm$0.10 & 3.14$\pm$0.02 & 3.72$\pm$0.14 & 5.8$\pm$0.7 & 2.36$\pm$0.06 & 2.18$\pm$0.01 & 2.86$\pm$0.15 & 5.1$\pm$0.5   \\
PBE0$^{a,b}$    &  -            & 3.13$\pm$0.02 & 3.78$\pm$0.08 & 6.2$\pm$0.6 &  -            & 2.21$\pm$0.01 & 2.90$\pm$0.06 & 5.2$\pm$0.3   \\
PBE0+TS-vdW(SC) & 2.52$\pm$0.17 & 3.16$\pm$0.02 & 3.73$\pm$0.15 & 6.3$\pm$0.8 & 2.28$\pm$0.17 & 2.18$\pm$0.03 & 2.90$\pm$0.11 & 5.5$\pm$0.2   \\
Expt.$^c$       & -             & 3.16$\pm$0.11 & -             & 6.9$\pm$1.0 &  -            & 2.19$\pm$0.16 & -             & 6.0$\pm$1.1   \\
Expt.$^d$       & -             & 3.11$\pm$0.03 & -             & 6.4$\pm$1.0 &  -            & -             & -             & -             \\
\hline\hline
\end{tabular}
\end{center}
$^a$ PBE0 results at  380 K from Ref. [\onlinecite{Zhang_2013}]. \\
$^b$ PBE0 results at  380 K from Ref. [\onlinecite{Gaiduk_2014}]. \\
$^c$ Empirical potential structure refinement (EPSR) of neutron diffraction data (0.67 M NaCl solutions) of Ref.~[\onlinecite{Mancinelli_2007}]. \\
$^d$ X-ray scattering experiment of Ref. [\onlinecite{Dang_2006}]. \\
\end{table*}

We begin our investigation of the aqueous chloride ion solution by considering the chloride--oxygen and chloride--hydrogen RDFs, $g_{\rm Cl-O}(r)$ and $g_{\rm Cl-H}(r)$, as shown in Fig.~\ref{rdf1}.
The numerical values for the intensities ($h_{\rm Cl-O}^{\rm max}$ and $h_{\rm Cl-H}^{\rm max}$) and positions ($r_{\rm Cl-O}^{\rm max}$ and $r_{\rm Cl-H}^{\rm max}$) of the first maximum, the positions ($r_{\rm Cl-O}^{\rm min}$ and $r_{\rm Cl-H}^{\rm min}$) of the first minimum, as well as the corresponding coordination numbers ($n_{\rm Cl-O}$ and $n_{\rm Cl-H}$) computed from these RDFs are provided in Table~\ref{table1} along with the available experimental findings.
Here, the $n_{\rm Cl-O}$ and $n_{\rm Cl-H}$ associated with the first coordination shell were computed by integrating each respective $g(r)$ in Fig.~\ref{rdf1} up to the first minimum \textit{via} $n_{\rm \alpha-\beta}=4\pi\rho\int dr\,g_{\rm \alpha-\beta}(r)$, wherein $\rho$ is the number density of the respective atom.

The first well-pronounced feature to note in Fig.~\ref{rdf1} is the sharp first peak in the $g_{\rm Cl-O}(r)$ corresponding to the first solvation shell, which is followed by a lower and broader second peak representing the second coordination shell.
In general, all of the AIMD trajectories considered here (generated using the four different XC functionals discussed above) yield consistent RDFs, with each predicting that the position of the first maximum is located at $r_{\rm Cl-O}^{\rm max} = 3.14 \pm 0.02$ \AA\ ($r_{\rm Cl-H}^{\rm max} = 2.17 \pm 0.02$ \AA).
However, there is a noticeable trend in that PBE0 and the two vdW-inclusive XC functionals predict values of $r_{\rm Cl-O}^{\rm max}$ and $r_{\rm Cl-H}^{\rm max}$ that are larger by 0.03--0.05 \AA\ in comparison to PBE.
In this regard, all of these values obtained from our simulations are in fairly good agreement with the experimental scattering values of 3.11--3.16 \AA\ for $r_{\rm Cl-O}^{\rm max}$ and 2.19 \AA\ for $r_{\rm Cl-H}^{\rm max}$ (see Table~\ref{table1}).
Furthermore, we note that our findings for the $r_{\rm Cl-O}^{\rm max}$ and $r_{\rm Cl-H}^{\rm max}$ quantities at the PBE0 level of theory are also in relatively good agreement with the analogous data provided by the recent work of Zhang \textit{et al.},~\cite{Zhang_2013} which studied the solvation structure of Cl$^-$ in water using the PBE0 XC functional (see Table~\ref{table1}).
With deviations of 0.01 \AA\ and 0.03 \AA, respectively, these small differences may be attributed to the $\sim$ 50 K temperature difference between these two simulations.

Since the first peaks of the $g_{\rm Cl-O}(r)$ and $g_{\rm Cl-H}(r)$ are useful indicators of the relative structuring of the solvation shell surrounding a given Cl$^-$ ion, the dependence of these quantities on the underlying XC potential deserves further comment.
For one, we note that the inclusion of vdW/dispersion interactions leads to a reduction in the structure of the first solvation shell---an effect that manifests as a simultaneous reduction in the intensities coupled with an increase in the positions of the first maxima in both $g_{\rm Cl-O}(r)$ and $g_{\rm Cl-H}(r)$. 
At the PBE (PBE0) level of theory, the inclusion of non-local vdW/dispersion interactions leads to significant reductions in $h_{\rm Cl-O}^{\rm max}$ and $h_{\rm Cl-H}^{\rm max}$ of 0.14 (0.18) and 0.26 (0.08), respectively.
Interestingly, the trend observed here indicates that the vdW/dispersion correction has a similar effect on $h_{\rm Cl-O}^{\rm max}$ when used in conjunction with either a GGA (PBE) or a hybrid (PBE0) XC functional, yet the same correction clearly has a more significant effect on $h_{\rm Cl-H}^{\rm max}$ when used in conjunction with the PBE-GGA functional.
Since the inclusion of exact exchange at the PBE0 level of theory weakens the hydrogen bond strength with respect to PBE~\cite{Distasio2014,santra_prl_2011,santra_jcp_2009,Zhang_2011}, this directly affects the interactions between the Cl$^-$ ion and the neighboring hydrogen atoms in the predominantly hydrogen-bonded first solvation shell.
Therefore, $h_{\rm Cl-H}^{\rm max}$ is significantly reduced with PBE0 compared to PBE (by 0.15), whereas the effect of PBE0 on $h_{\rm Cl-O}^{\rm max}$ is only marginal (by 0.03).
In other words, PBE0 significantly softens the first peak in the $g_{\rm Cl-H}(r)$, but has little effect on the first peak in the $g_{\rm Cl-O}(r)$.
Hence, the effects of non-local vdW/dispersion interactions on $h_{\rm Cl-H}^{\rm max}$ are significantly different when used in conjunction with PBE and PBE0, while their effects on $h_{\rm Cl-O}^{\rm max}$ are quite similar.
When compared to PBE, the collective effects of exact exchange and vdW/dispersion interactions, \textit{i.e.}, as accomplished with the use of the PBE0+TS-vdW(SC) XC functional, reduce $h_{\rm Cl-O}^{\rm max}$ and $h_{\rm Cl-H}^{\rm max}$ by 0.15 and 0.23, accompanied by small increases in $r_{\rm Cl-O}^{\rm max}$ and $r_{\rm Cl-H}^{\rm max}$ of 0.05 \AA\ and 0.04 \AA, respectively.
From these two observations, it is clear that the inclusion of exact exchange and non-local vdW/dispersion interactions is effectively \textit{weakening} the interaction between the Cl$^-$ ion and the surrounding water molecules in the first solvation shell.

The effects of vdW/dispersion interactions are also clearly evident in the interstitial region between the first and second coordination shells in both $g_{\rm Cl-O}(r)$ and $g_{\rm Cl-H}(r)$.
Here, the attractive vdW/dispersion force enhances the interaction strength between the Cl$^-$ ion and interstitial water molecules and favors more distorted and entropically-driven geometrical configurations. 
Hence, we observed a net increase in the population of water molecules in the interstitial region in both $g_{\rm Cl-O}(r)$ and $g_{\rm Cl-H}(r)$ with the vdW-inclusive functionals considered in this work (see Fig.~\ref{rdf1}).

\begin{figure}
\includegraphics[width=8.7cm]{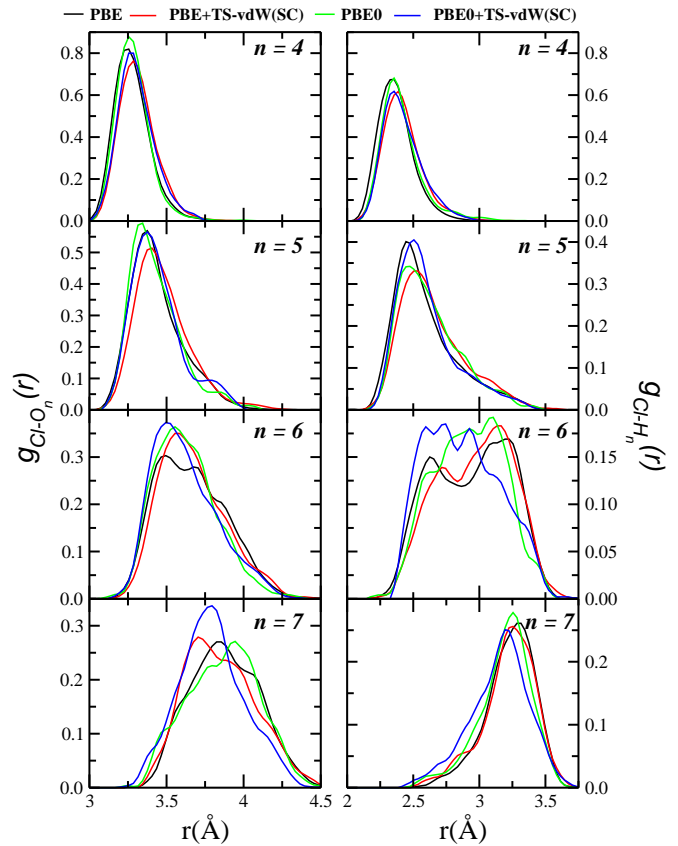}
\caption{Contributions from the 4th ($n=4$), 5th ($n=5$), 6th ($n=6$), and 7th ($n=7$) nearest neighbors to the Cl-O (left panel) and Cl-H (right panel) radial distribution functions, $g_{\rm Cl-O}(r)$ and $g_{\rm Cl-H}(r)$, of the aqueous chloride ion solution obtained from theory via the DFT-based AIMD simulations performed in this work.} 
\label{fig:pRDF}
\end{figure}

This simultaneous weakening and strengthening of the interactions between the Cl$^-$ ion and its surrounding water molecules as obtained with vdW-inclusive hybrid DFT can be further understood by considering a partial decomposition of the RDFs.
In Fig.~\ref{fig:pRDF}, the contributions to the $g_{\rm Cl-O}(r)$ and $g_{\rm Cl-H}(r)$ arising from the 4th, 5th, 6th, and 7th neighboring water molecules are plotted for each of the XC functionals employed in this work.
At the PBE0+TS-vdW(SC) level of theory, the first five oxygen and hydrogen atoms (with only the 4th and 5th neighbors shown for clarity) were found to be further from the Cl$^-$ ion on average, as compared to the corresponding distances obtained with PBE, indicating a weakening of the interactions between the Cl$^-$ ion and \textit{these} surrounding water molecules.
However, the exact opposite trend was found for the 6th and 7th neighboring oxygen and hydrogen atoms, in which these neighbors were found to be much closer to the Cl$^-$ ion in the case of PBE0+TS-vdW(SC) when compared to PBE, indicating a strengthening of the interactions with the Cl$^-$ ion and \textit{these} surrounding water molecules.
As a result, the collective effects of exact exchange and vdW/dispersion interactions tend to weaken the interactions between the Cl$^-$ ion and the first five neighbors while simultaneously strengthening the interactions with the water molecules beyond the 6th neighbors, which results in a larger overlap between the first and second coordination shells and a net reduction in the structure of the solvation shell surrounding the Cl$^-$ ion.

\begin{figure}
\includegraphics[width=8.0cm]{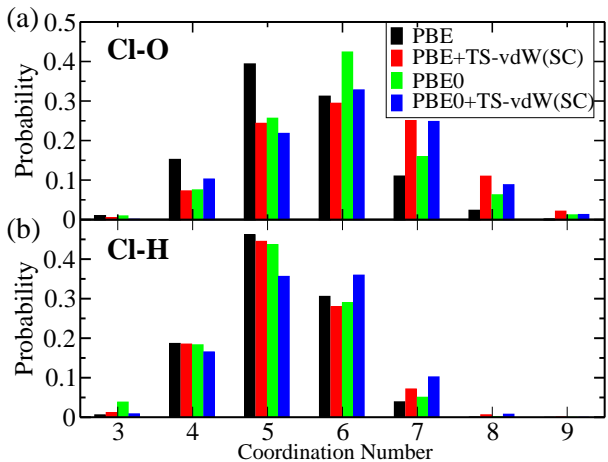}
\caption{Probability distributions of the (a) Cl-O and (b) Cl-H coordination numbers, $n_{\rm Cl-O}$ and $n_{\rm Cl-H}$, calculated by integrating $g_{\rm Cl-O}(r)$ and $g_{\rm Cl-H}(r)$ up to their respective first minimum.}
\label{fig:coord}
\end{figure}

\begin{figure*}[ht]
\vspace {0.5cm}
\includegraphics[width=14.0cm]{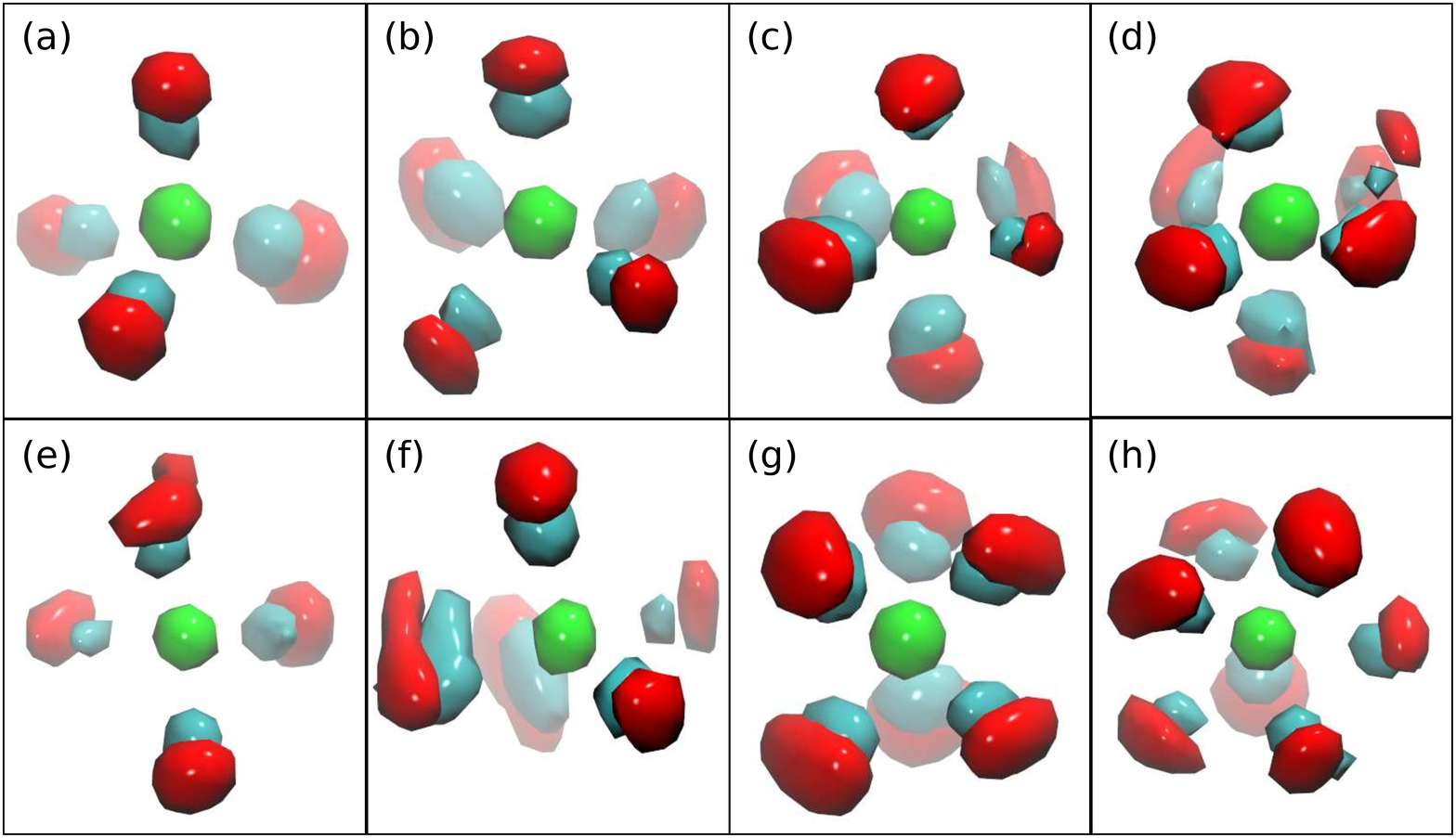}
\caption{Spatial density functions (SDF) of the first solvation shell surrounding the aqueous Cl$^-$ ion characterized by 4-, 5-, 6-, and 7-fold coordination numbers, respectively (shown from left to right). The top ((a)--(d)) and bottom ((e)--(h)) panels correspond to structures (configurations) obtained from PBE and PBE0+TS-vdW(SC) AIMD simulations, respectively. The Cl$^-$ ion, oxygen atoms, and hydrogen atoms are depicted by green, red, and cyan spheres, respectively.}
\label{fig:SDF}
\end{figure*}

Furthermore, the collective effects of exact exchange and vdW/dispersion interactions also influence the positions of the first minima in the $g_{\rm Cl-O}(r)$ and $g_{\rm Cl-H}(r)$, quantities which are typically used to indicate the points of separation between the first and second coordination shells.
In this regard, the positions of the first minima in these RDFs, \textit{i.e.}, $r_{\rm Cl-O}^{\rm min}$ and $r_{\rm Cl-H}^{\rm min}$, increase by 0.09 and 0.02 \AA, respectively, with PBE0+TS-vdW(SC) in comparison to PBE (see Table~\ref{table1}).
This finding has an important immediate consequence, in that the mean coordination numbers obtained by integrating the first peak of these RDFs up to their first respective minima vary considerably among the XC functionals considered herein.
Here, we found that the mean Cl-O coordination number, $n_{\rm Cl-O}$, increases by 0.8 with both PBE+TS-vdW(SC) and PBE0+TS-vdW(SC) in comparison to PBE (see Table~\ref{table1}) and the resultant mean value of $n_{\rm Cl-O} = 6.3$ agrees reasonably well with recent experimental findings that report coordination numbers between 6.4--6.9 in $<2$M aqueous NaCl solutions.~\cite{Mancinelli_2007,Dang_2006}
For the mean Cl-H coordination number, PBE0+TS-vdW(SC) simulations yield a value of $n_{\rm Cl-H}=5.5$, which is the largest value among all of the XC functionals considered in this work (all of which predict a consistent mean value of $\approx$ 5.2), and is in better agreement with the experimental EPSR value of 6.0$\pm$1.1.~\cite{Mancinelli_2007}
We note in passing that these estimates for the mean Cl-O and Cl-H coordination numbers are consistent with previous AIMD simulations performed with GGA and hybrid functionals.~\cite{Ikeda_2003,Heuft_2003,Tongraar_2003,Mallik_2008,Scipioni_2009,Guardia_2009,Calemana_2011,Bankura_2013,Ge_2013,Zhang_2013,Gaiduk_2014}

In order to further elucidate the instantaneous variations in the geometry of the first coordination shell surrounding the aqueous Cl$^-$ ion, we now analyze the probability distributions of the $n_{\rm Cl-O}$ and $n_{\rm Cl-H}$ coordination numbers (see Fig.~\ref{fig:coord}).
From this analysis, we found that most ($\approx$ 40--50\%) of snapshots at the PBE level of theory were characterized by 5-fold coordination for both $n_{\rm Cl-O}$ and $n_{\rm Cl-H}$ (see Fig.~\ref{fig:coord}), indicating that the Cl$^-$ ion is predominantly solvated by five water molecules, each of which are pointing one hydrogen atom toward the Cl$^-$ ion.
At the PBE0+TS-vdW(SC) level of theory, we instead found that most ($\approx$ 35\%) of snapshots can be characterized by 6-fold Cl-O coordination and that both 5- and 6-fold Cl-H coordinations were found to be equally most probable ($\approx$ 35\% each).
Qualitatively similar observations can also be made with the PBE0 and PBE+TS-vdW(SC) XC functionals, in which $n_{\rm Cl-O}$ is dominated by 6-fold coordination, whereas $n_{\rm Cl-H}$ is dominated by 5-fold coordination.
This apparent mismatch between the most dominant coordination number in the $n_{\rm Cl-O}$ and $n_{\rm Cl-H}$ probability distributions is reflective of the presence of more disordered hydrogen-bond structures in the first solvation shell.

To graphically depict the differential coordination numbers found in the aqueous Cl$^-$ ion solution, we have computed the spatial density functions (SDF) within the first solvation shell of the Cl$^-$ ion based on structures (configurations) obtained at the PBE and PBE0+TS-vdW(SC) levels of theory.
In Fig.~\ref{fig:SDF}, we have plotted SDFs for configurations in which the Cl$^-$ ion was coordinated with 4, 5, 6, and 7 water molecules, respectively, averaged over all structures characterized by these coordination numbers.
For the case of 4-fold coordination, we found that distorted tetrahedral structures are dominant over square planar structures in both XC functionals (see Fig.~\ref{fig:SDF}(a) and Fig.~\ref{fig:SDF}(e)).
In both XC functionals, the 5-fold coordinated aqueous Cl$^-$ ion complexes are predominantly comprised of distorted square pyramidal structures instead of trigonal bipyramidal structures; however, the SDF of the oxygen and hydrogen atoms obtained with PBE0+TS-vdW(SC) (see Fig.~\ref{fig:SDF}(f)) were found to be more delocalized than that obtained with PBE (see Fig.~\ref{fig:SDF}(b)). 
For the case of 6-fold coordination, the predominant form found with PBE resembles an octahedral arrangement (see Fig.~\ref{fig:SDF}(c)), whereas the trigonal prism structure was more representative of the PBE0+TS-vdW(SC) configurations (see Fig.~\ref{fig:SDF}(g)). 
The 7-fold coordination case again significantly differs among these distinct XC functionals; here the predominant shape found at the PBE level of theory resembles a capped octahedron (see Fig.~\ref{fig:SDF}(d)), while the more common pentagonal bipyramidal arrangement dominates at the PBE0+TS-vdW(SC) level of theory (see Fig.~\ref{fig:SDF}(h)).
In all of the SDF plots in Fig..~\ref{fig:SDF}, the oxygen and hydrogen atoms were found to be more delocalized on average in the PBE0+TS-vdW(SC) configurations.
This observation, in conjunction with the earlier finding that a majority of the configurations posses 6- and 7-fold Cl-O coordination, again exemplifies the fact that the first solvation shell surrounding the Cl$^-$ ion is significantly more disordered at the vdW-inclusive hybrid PBE0+TS-vdW(SC) level of theory.

\begin{figure}
\vspace {0.5cm}
\includegraphics[width=8.0cm]{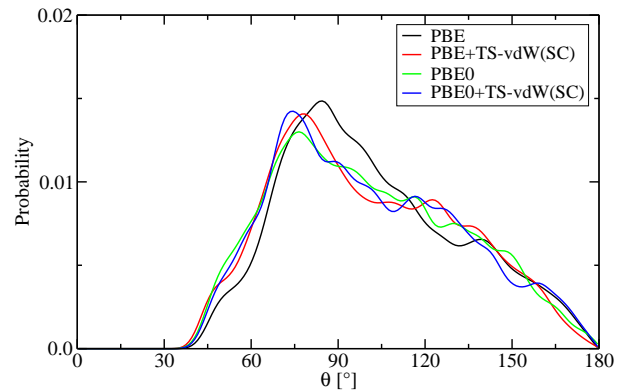}
\caption {Angular distribution functions, O-Cl-O, for each of the XC functionals employed herein computed by considering only the oxygen atoms which reside within the first coordination shell surrounding the Cl$^-$ ion (as defined by $r_{\rm Cl-O}^{\rm min}$, the position of the first minimum in the $g_{\rm Cl-O}(r)$).}
\label{fig:ang2}
\end{figure}

In addition to the detailed coordination structures considered above, the aforementioned disorder in the first coordination shell surrounding the Cl$^-$ ion can also be characterized by analyzing the O-Cl-O angular distribution function.
In this regard, O-Cl-O angular distribution functions were computed for each of the XC functionals employed herein by considering only the oxygen atoms which reside within the first coordination shell surrounding the Cl$^-$ ion (defined by the radial cutoff distance given by $r_{\rm Cl-O}^{\rm min}$).
As seen in Fig.~\ref{fig:ang2}, the PBE XC functional predicts a maximum at $\theta\approx90^{\circ}$, a finding which is consistent with the fact that a majority of the configurations at the PBE level are characterized by 5-fold Cl-O coordination (see Fig.~\ref{fig:coord}) and predominantly square pyramidal structures (see Fig.~\ref{fig:SDF}(b)).
On the other hand, PBE0 and both vdW-inclusive XC functionals predict maxima at $\theta\approx75^{\circ}$, which is consistent with the fact that a majority of the configurations at these levels of theory are characterized by 6-fold Cl-O coordination (see Fig.~\ref{fig:coord}) and predominantly distorted trigonal prism structures (see Fig.~\ref{fig:SDF}(g)).
Hence, this analysis of the O-Cl-O angular distribution functions is again strongly indicative of the net increase in the amount of structural disorder present in the first solvation shell surrounding the Cl$^-$ ion when both exact exchange and non-local vdW/dispersion interactions are accounted for in the underlying XC potential.

\noindent

\subsection{Electronic Structure of the Aqueous Cl$^-$ Ion \label{sec:electronic}}

Beyond the solvation structures of aqueous ionic solutions, which can be directly compared to diffraction experiments, the electronic properties of aqueous ionic solutions, such as the band structure of the solution and the energy levels of the solvated ion, are also important quantities that can be probed by photoemission (PES) and X-ray absorption spectroscopy (XAS).
In this regard, a recent PES experiment~\cite{Winter_2006} has measured electron binding energies in aqueous chloride ion solutions, in which the position of the highest occupied molecular orbital (HOMO) of the solvated chloride ion was found to be $\delta E$ = 1.25--1.50 eV above the valence band maximum (VBM) of liquid water.
In order to theoretically study the electronic structure properties of the aqueous Cl$^-$ ion solution, we have performed DFT-based ground state wavefunction optimizations based on equilibrated structures (configurations) from various AIMD trajectories.
In particular, we have also adopted several different XC approximations, ranging from semi-local GGAs (PBE) to hybrid functionals (PBE0 and BHLYP) in these band structure calculations.

We begin this analysis by first considering the band gaps (HOMO-LUMO gaps) of liquid water and the aqueous Cl$^-$ ion (see Table~\ref{table2}).
Here, it is well-known that the accuracy of band structure calculations depends crucially on the XC approximation employed in the quantum mechanical description of the electrons.
In this regard, the inherent delocalization error and the resulting incorrect convex behavior characteristic of semi-local GGA functional approximations leads to large underestimates of band gaps and incorrect predictions for other features in the band structure.~\cite{Yang_Science}
Consistent with previous DFT calculations of liquid water employing the PBE functional,~\cite{Zhang_2013} the estimated band gap was found to be $\approx$ 4.4 eV in this work, which is underestimated by $\approx$ 4.3 eV from the best experimental estimates.~\cite{Winter_2006}
Incorporation of a fraction of exact exchange ($E_{\rm xx}$) is an established approach for remedying this problem to some extent,~\cite{Adriaanse_2012,Liu_2015,Cheng_2014,Pham_2014,Viktor_2013} yielding band gaps of approximately 7.0 and 9.7 eV for liquid water using the PBE0 (25\% $E_{\rm xx}$) and BHLYP (50\% $E_{\rm xx}$) hybrid XC functionals, respectively (see Table~\ref{table2}).
Hence, the inclusion of $E_{\rm xx}$ reduces the delocalization (self-interaction) error and leads to a marked improvement in the predicted band gaps of liquid water.
In this regard, it should also be noted that the computed band gaps are relatively insensitive to the molecular structures (configurations) obtained from AIMD trajectories generated by various XC functional approximations.~\cite{Zhang_2013}

\begin{table}[t]
\caption{Electronic structure (ES) properties computed with the PBE, PBE0, and BHLYP XC functionals based on structures (configurations) generated with AIMD simulations at the PBE, PBE+TS-vdW(SC), PBE0, and PBE0+TS-vdW(SC) levels of theory. $E_{\rm g}^{\rm H_2O}$ and $E_{\rm g}^{\rm sol}$ are the computed band gaps (in eV) for liquid water and the aqueous Cl${^-}$ ion solution, respectively. $\delta$E is the energetic difference between the HOMO of the Cl$^-$ ion and the valence band maximum (VBM) of liquid water.}
\label{table2}
\begin{center}
\begin{tabular}{c|c|c|c}
\hline\hline
ES Level//AIMD Level  & $E_{\rm g}^{\rm H_2O}$ & $E_{\rm g}^{\rm sol}$ & $\delta E$  \\
\hline
PBE//PBE               & -                      & 4.22                  & -0.24       \\
PBE0//PBE               & -                      & 6.68                  &  0.13       \\
BHLYP//PBE               & -                      & 8.81                  &  0.72       \\
\hline
PBE//PBE+TS-vdW(SC)    & -                      & 4.16                  & -0.11       \\
PBE0//PBE+TS-vdW(SC)    & -                      & 6.53                  &  0.27       \\
BHLYP//PBE+TS-vdW(SC)    & -                      & 8.61                  &  0.88       \\
\hline
PBE//PBE0              & -                      & 4.19                  & -0.06       \\
PBE0//PBE0              & -                      & 6.55                  &  0.35       \\
BHLYP//PBE0              & -                      & 8.66                  &  0.95       \\
\hline
PBE//PBE0+TS-vdW(SC)   & 4.42                   & 4.20                  & -0.06       \\
PBE0//PBE0+TS-vdW(SC)   & 7.04                   & 6.57                  &  0.34       \\
BHLYP//PBE0+TS-vdW(SC)   & 9.70                   & 8.62                  &  0.92       \\
\hline\hline
\end{tabular}
\end{center}
\end{table}

We now focus our attention on the electronic properties of the aqueous Cl$^-$ ion solution.
In particular, $\delta E$, which is defined as the relative energetic difference between the solvated Cl$^-$ ion and the VBM of liquid water will play a key role in our analysis.
In this regard, it was previously found that the energy level of the solvated Cl$^-$ ion is quite sensitive to both the underlying molecular structure as well as the choice of the electronic structure method employed during ground state wavefunction optimization.~\cite{Zhang_2013}
Therefore, a systematic study that differentiates these two effects is necessary and will be addressed in this work.

\begin{figure}[ht]
\includegraphics[width=8.0cm]{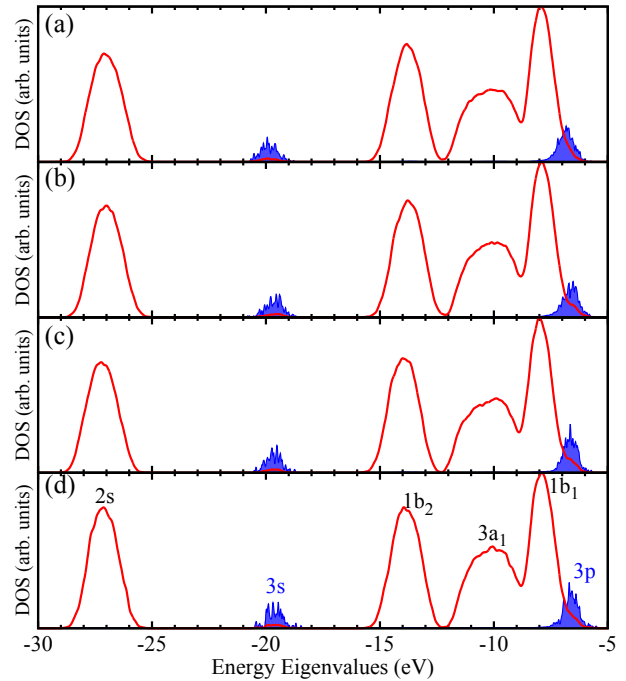}
\caption {Total density of states (DOS in red) and real-space projected density of states (PDOS in blue) for the aqueous Cl$^-$ ion solution computed using the hybrid PBE0 XC functional based on structures (configurations) generated from AIMD trajectories at the (a) PBE, (b) PBE+TS-vdW(SC), (c) PBE0, and (d) PBE0+TS-vdW(SC) levels of theory. The DOS and PDOS above were obtained from electronic structure calculations on 200 configurations taken from each AIMD trajectory.}.
\label{fig:all-pbe0}
\end{figure}

To accomplish this goal, we first performed band structure calculations with the hybrid PBE0 XC functional based on structures (configurations) generated with AIMD simulations at the PBE, PBE+TS-vdW(SC), PBE0, and PBE0+TS-vdW(SC) levels of theory. 
With a self-consistent solution to the non-linear Kohn-Sham (KS) equations, we projected the optimized ground-state KS eigenfunctions onto the solvated Cl$^-$ ion with a radius of $R=1.5$ \AA.~\cite{Swartz_2013}
The resulting total density of states (DOS) and real-space projected DOS (PDOS) computed at the PBE0 level of theory are plotted in Fig.~\ref{fig:all-pbe0}, from which it is evident that the total DOS of the aqueous Cl$^-$ ion solution is dominated by the features belonging to the valence electrons of liquid water.
In this regard, the oxygen $2p$ states are located well above ($>10$ eV) the oxygen semi-core $2s$ states, and can be further decomposed into three features, corresponding to the $1b_2$, $3a_1$, and $1b_1$ bands representing the covalently bonded and non-bonded (lone pair) electrons.
As far as the PDOS of the solvated Cl$^-$ ion is concerned, the $3s$ states of the Cl$^-$ ion were found to be between the $2s$ and $2p$ bands of liquid water while the $3p$ states of the Cl$^-$ ion were found to be located at the edge of the VBM of liquid water---a picture which is consistent with the PES measurements in aqueous Cl$^-$ ionic solutions.~\cite{Winter_2006}
Moreover, it can be seen that the total DOS computed based on molecular configurations generated from different AIMD trajectories remain very similar in terms of the peak positions and spectral distributions.
This finding is indicative of the fact that the band structure of liquid water is rather insensitive to the underlying molecular configurations (\textit{i.e.}, to whether or not $E_{\rm xx}$ and/or non-local vdW/dispersion interactions are accounted for in the underlying XC potential).

In sharp contrast, the relative position of the orbital levels associated with the solvated Cl$^-$ ion with respect to the VBM of liquid water shows a much stronger dependence on the underlying molecular configurations.
As plotted in Fig.~\ref{fig:all-pbe0}, the PDOS computed at the PBE0 level of theory based on molecular configurations generated from the PBE-AIMD trajectory is mostly located below the VBM of liquid water, yielding a value of $\delta E = 0.13$ eV (see Table~\ref{table2}).
With exact exchange and/or non-local vdW/dispersion interactions accounted for in the underlying potential to generate the AIMD trajectories, the distribution of the PDOS becomes more prominantly separated from the VBM of liquid water.
As a result, the computed $\delta E$ increases in the direction of the experimental value of $\delta E =$ 1.25--1.50 eV as shown in Table~\ref{table2}, a result that we attribute to the weakened interaction between the Cl$^-$ ion and its surrounding water molecules with the use of vdW-inclusive hybrid XC functionals.
As discussed above in Sec.~\ref{sec:solvation}, the collective effects of $E_{\rm xx}$ and non-local vdW/dispersion interactions lead to a net decrease in the strength of the hydrogen bonds existing between the lone pair electrons of the Cl$^-$ ion and the water molecules residing in the first solvation shell.
With the positions of the first maximum in $g_{\rm Cl-O}(r)$ and $g_{\rm Cl-H}(r)$, \textit{i.e.}, $r_{\rm Cl-O}^{\rm max}$ and $r_{\rm Cl-H}^{\rm max}$, slightly increased, hybridization between the Cl$^-$ $3p$ and the oxygen $2p$ orbitals (located at the edge of the liquid water valence band) becomes less favorable.
We note in passing that this explanation is consistent with the increased $\delta E$ depicted in Fig.~\ref{fig:all-pbe0}.

\begin{figure}[ht]
\includegraphics[width=8.0cm]{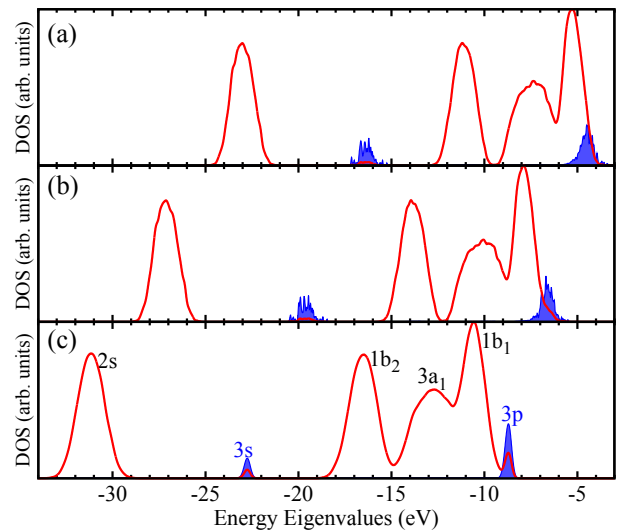}
\caption {Total density of states (DOS in red) and real-space projected density of states (PDOS in blue) for the aqueous Cl$^-$ ion solution computed using the (a) PBE, (b) PBE0, and (c) BHLYP XC functionals based on structures (configurations) generated from an AIMD trajectory at the PBE0+TS-vdW(SC) level of theory. The DOS and PDOS above were obtained from electronic structure calculations on 200 configurations taken from the AIMD trajectory.}
\label{fig:pbe0-vdw}
\end{figure}

From the above discussion, one can see that an accurate prediction of the electronic properties of the solvated Cl$^-$ ion crucially depends on the description of the interaction between the $p$ electrons of Cl and O.
In this regard, the more accurate prediction of the hydrogen bond tends to discourage hybridization among these $p$ orbitals, thereby increasing $\delta E$ towards the experimental value.
This trend is even more pronounced when different XC approximations are utilized in the theoretical treatment of the electronic structure.
As shown in Fig.~\ref{fig:pbe0-vdw}, we now present the DOS and PDOS of the aqueous Cl$^-$ ion solution based on ground-state wavefunction optimizations performed at the PBE, PBE0, and BHLYP levels of theory.
In this case, the underlying molecular structures or configurations are kept fixed and were taken from an AIMD trajectory employing the vdW-inclusive hybrid PBE0+TS-vdW(SC) XC potential.
This choice is motivated by our recent study in which we found that the PBE0+TS-vdW(SC) XC potential yielded a very accurate prediction of the microscopic structure of ambient liquid water that is in quantitative agreement with respect to the available scattering experiment data.~\cite{Distasio2014}

\begin{figure}[ht]
\includegraphics[width=8.0cm]{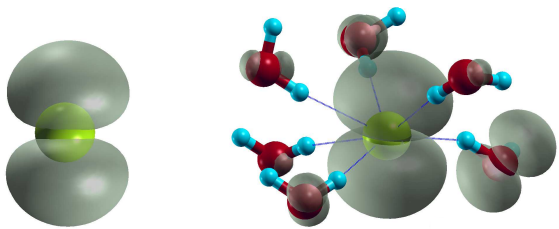}
\caption {The triply degenerate gas-phase (left) and condensed-phase (right) $3p$ orbitals for the Cl$^-$ ion.}
\label{Fig:orbital}
\end{figure}

As clearly seen from Fig.~\ref{fig:pbe0-vdw}, the PDOS of the solvated Cl$^-$ ion is significantly affected by the choice of DFT functional employed in the electronic structure calculation. 
At the semi-local PBE-GGA level of theory, the computed value of $\delta E = -0.06$ eV is negative, which is opposite in sign compared to the experimental value; however, as the fraction of $E_{\rm xx}$ is increased from 25\% (PBE0) to 50\% (BHLYP), the $\delta E$ steadily increases to 0.34 eV and 0.92 eV, respectively, and is rapidly approaching the experimental value.
Again, we attribute this trend to the removal of the deleterious self-interaction error that is present in semi-local XC functionals \textit{via} the admixture of exact exchange in the hybrid PBE0 and BHLYP XC functionals.
As such, the electronic structure predicted by semi-local GGA functionals (such as PBE) artificially favors the interaction of the Cl$^-$ ion and its surrounding water molecules by facilitating a greater extent of $p$ orbital hybridization.
This effect is greatly reduced at the hybrid DFT level of theory---at the BHLYP level, the eigenfunctions of the Cl$^-$ ion HOMO were found to have a non-negligible amplitude located on water molecules in the first coordination shell in only $\approx$ 5\% of the configurations.
However, we note in passing that this fraction increases to over 70\% when the semi-local PBE-GGA XC functional is used instead (see Fig.~\ref{Fig:orbital}).

These findings reported above are consistent with the previous study by Zhang \textit{et al.}~\cite{Zhang_2013} in which it was found that the use of hybrid XC functionals in generating both the electronic and molecular structure was essential for a qualitatively correct prediction of $\delta E$.
The current study further indicates that accounting for non-local vdW/dispersion interactions in the generation of the underlying molecular structure is also important for obtaining a quantitatively correct theoretical prediction of this fundamental quantity in aqueous ionic solutions.
By admixing 25\% of $E_{\rm xx}$, the hybrid PBE0 XC functional has successfully generated an electronic structure that is qualitatively consistent with the available PES experiments.
However, 25\% of $E_{\rm xx}$ might not be sufficient to completely eliminate the effects of self-interaction error on $\delta E$,~\cite{Private} which can serve as an explanation for the remaining discrepancy that exists between theory and experiment.
In addition, it should also be noted that an explicit treatment of nuclear quantum effects (NQE), as provided by the Feynman discretized path integral technique, is expected to further decrease the strength of the interaction between the Cl$^-$ ion and water---as a result, a further increase in $\delta E$ towards the experimental value is anticipated with such a theoretical treatment of this fundamental aqueous ionic solution.

\section{Conclusions}

In this work, we have systematically studied the solvation and electronic structure of the aqueous chloride ion solution using DFT-based AIMD which incorporates exact exchange ($E_{\rm xx}$) and non-local vdW/dispersion interactions in the underlying XC potential.
From an analysis of $g_{\rm Cl-O}(r)$ and $g_{\rm Cl-H}(r)$, we find that the inclusion of exact exchange and non-local vdW/dispersion interactions \textit{via} the use of the PBE0+TS-vdW(SC) XC functional effectively \textit{weakens} the interactions between the Cl$^-$ ion and the surrounding water molecules residing in the first solvation shell and yields a mean Cl-O coordination number, $n_{\rm Cl-O}=6.3\pm0.8$, that is in excellent agreement with the available experimental data.
Moreover, we found that most configurations at the vdW-inclusive hybrid PBE0+TS-vdW(SC) level of theory can be characterized by 6-fold Cl-O coordination and are predominantly comprised of distorted trigonal prism structures---findings which are strongly indicative of a significantly disordered first solvation shell surrounding the Cl$^-$ ion.

By performing a systematic series of band structure calculations on molecular structures (configurations) generated from AIMD simulations with various differing XC potentials, we were able to selectively isolate the effects arising from the underlying molecular structure as well as the choice of the electronic structure method on the energy levels of the solvated Cl$^-$ ion relative to the VBM of liquid water ($\delta E$).
In doing so, we found that while the band structure of liquid water is rather insensitive to the underlying molecular configurations, the relative positions of the orbital levels associated with the solvated Cl$^-$ ion with respect to the VBM of liquid water strongly depend on the underlying molecular configurations.
Here, the effective weakening of the hydrogen bonds existing between the lone pair electrons of the Cl$^-$ ion and the water molecules residing in the first solvation shell due to the collective effects of $E_{\rm xx}$ and non-local vdW/dispersion interactions leads to a decrease in the degree of hybridization between the Cl$^-$ $3p$ and the oxygen $2p$ orbitals located at the edge of the liquid water valence band.
In the same breath, the PDOS of the solvated Cl$^-$ ion is also significantly affected by the choice of DFT functional employed in the electronic structure calculation; as the fraction of $E_{\rm xx}$ was increased from 25\% (PBE0) to 50\% (BHLYP), $\delta E$ steadily increased from 0.34 eV to 0.92 eV, thereby rapidly approaching the experimental value of 1.25--1.50 eV.

As a final set of remarks, we note that the quantitative description of the microscopic structure of the aqueous Cl$^-$ ion solution may change by further refinement of the underlying XC functional approximation, which may be accomplished by reducing the self-interaction error \textit{via} fine-tuning the exchange correction~\cite{Skone2014}, and/or including a better description of vdW/dispersion interactions \textit{via} the inclusion of beyond-pairwise interactions, as provided, for example, by the recently proposed many-body dispersion (MBD) scheme~\cite{tkatchenko_prl_2012,distasio_pnas_2012,ambrosetti_jcp_2014,distasio_jpcm_2014}.
As mentioned above, structural refinements are also expected to result from a proper treatment of nuclear quantum effects \textit{via} the Feynman discretized path-integral approach, which is expected to further weaken the interactions between the solvated Cl$^-$ ion and its aqueous surroundings; the use of such configurations in conjunction with a higher-level electronic structure theory, such as Hedin's GW self-energy approximation~\cite{GW1} and/or the random phase approximation (RPA)~\cite{RPA1,RPA2,RPA3}, should lead to even better agreement with PES experiments on the relative energetics among the solvated Cl$^-$ orbitals and the VBM of liquid water.

\begin{acknowledgments}
A.B. and M.K are grateful for partial financial support from the U.S. Department of Energy, Office of Basic Energy Sciences, SciDAC Award DE-FG02-12ER16333 and computational support from XSEDE Grant No. MCA93S020.
X.W. acknowledges support from the American Chemical Society Petroleum Research Fund (ACS PRF) under Grant No. 53482-DNI6.
R.D. and B.S. acknowledge support from the Scientific Discovery through Advanced Computing (SciDAC) program through the Department of Energy (DOE) under Grant Nos. DE-SC0005180 and DE-SC0008626.
This research used resources of the National Energy Research Scientific Computing Center, which is supported by the Office of Science of the U.S. Department of Energy under Contract No. DE-AC02-05CH11231.
This research used resources of the Argonne Leadership Computing Facility at Argonne National Laboratory, which is supported by the Office of Science of the U.S. Department of Energy under contract DE-AC02-06CH11357.
\end{acknowledgments}


\begin{thebibliography}{300}

\bibitem{Abbatt_2006} J. P. D. Abbatt, S. Benz, D. J. Cziczo, Z. Kanji, U. Lohmann, and O. M{\"{o}}hler, Science \textbf{313}, 1770 (2006).

\bibitem{Knipping_2000} E. M. Knipping, M. J. Lakin, K. L. Foster, P. Jungwirth, D. J. Tobias, R. B. Gerber, D. Dabdub, and B. J. Finlayson-Pitts, Science \textbf{288}, 301 (2000).

\bibitem{Spicer_1998} C. W. Spicer, E. G. Chapman, B. J. Finlayson-Pitts, R. A. Plastridge, J. M. Hubbe, J. D. Fast, and C. M. Berkowitz, Nature \textbf{394}, 353 (1998).

\bibitem{Beekman_2011} H. E. Beekman, H. G. M. Eggenkamp, and C. A. J. Appelo, Appl. Geochem. \textbf{26}, 257 (2011).

\bibitem{Cummings_1980} S. Cummings, J. E. Enderby, G. W. Neilson, J. R. Newsome, R. A. Howe, W. S. Howells, and A. K. Soper, Nature (London) \textbf{287}, 714 (1980).

\bibitem{Copestake_1985} A. P. Copestake, G. W. Neilson, and J. E. Enderby, J. Phys. C. \textbf{18}, 4211 (1985).

\bibitem{Yamagami_1995} M. Yamagami, H. Wakita, and T. Yamaguchi, J. Chem. Phys. \textbf{103}, 8174 (1995).

\bibitem{Megyes_2008} T. Megyes, B. Szabolcs, G. Tam{\'{a}}s, R. Tam{\'{a}}s, B. Imre, and S. P{\'{a}}l, J. Chem. Phys. \textbf{128}, 044501 (2008).

\bibitem{Tongraar_2010} A. Tongraar, J. T. Thienprasert, and S. Rujirawat, Phys. Chem. Chem. Phys. \textbf{12}, 10876 (2010).

\bibitem{Dang_2006} L. X. Dang, G. K. Schenter, V.-A. Glezakou, and J. L. Fulton, J. Phys. Chem. B \textbf{110}, 23644 (2006).

\bibitem{Bruni_2012} F. Bruni, S. Imberti, R. Mancinelli, and M. A. Ricci, J. Chem. Phys. {\bf 136}, 064520 (2012).

\bibitem{Mancinelli_2007} R. Mancinelli, A. Botti, F. Bruni, M. A. Ricci, and A. K. Soper, J. Phys. Chem. B {\bf 111}, 13570 (2007).

\bibitem{Mancinelli_2007_2} R. Mancinelli, A. Botti, F. Bruni, M. A. Ricci, and A. K. Soper, Phys. Chem. Chem. Phys. {\bf 9}, 2959 (2007).

\bibitem{Soper_2006} A. K. Soper and K. Weckstr{\"{o}}m, Biophys. Chem. {\bf 124}, 180 (2006).

\bibitem{Megyes_2006}T. Megyes, I. Bako, S. Balint, T.  Grosz, and T. Radnai, J. Mol. Liq. {\bf 129}, 63 (2006).

\bibitem{Botti_2004} A. Botti, F. Bruni, S. Imberti, M. A. Ricci, and A. K. Soper, J. Chem. Phys. {\bf 121}, 7840 (2004).

\bibitem{Delahay_1982} P. Delahay, Acc. Chem. Res. {\bf 15}, 40 (1982).

\bibitem{Ghosal_2005}S. Ghosal, J. C. Hemminger, H. Bluhm, B. S. Mun, E. L. D. Hebenstreit, G. Ketteler, D. F. Ogletree, F. G. Requejo and M. Salmeron, Science {\bf 307}, 563 (2005).

\bibitem{Kropman_2001}M. F. Kropman and H. J. Bakker, Science {\bf 291}, 2118 (2001).

\bibitem{Kropman_2002}M. Kropman, H. Nienhuys and H. J. Bakker, Phys. Rev. Lett. {\bf 88}, 077601 (2002).

\bibitem{Bakker_2005} H. Bakker, M. Kropman and A. Omta, J. Phys.: Condens. Matter {\bf 17}, S3215 (2005).

\bibitem{Skinner_2012} J. L. Skinner, P. A. Pieniazek and S. M. Gruenbaum, Acc. Chem. Res. {\bf 45}, 93 (2012).

\bibitem{Winter_2005} B. Winter, R. Weber, I. V. Hertel, M. Faubel, P. Jungwirth, E. C. Brown and S. E. Bradforth, J. Am. Chem. Soc. {\bf 127}, 7203 (2005).

\bibitem{Winter_2006}B. Winter, M. Faubel, I. V. Hertel, C. Pettenkofer, S. E. Bradforth, B. Jagoda-Cwiklik, L. Cwiklik and P. Jungwirth, J. Am. Chem. Soc. {\bf 128}, 3864 (2006).

\bibitem{Seidel_2011} R. Seidel, S. Thurmer and B. Winter, J. Phys. Chem. Lett. {\bf 2}, 633 (2011).

\bibitem{laasonen_jcp_1993} K. Laasonen, M. Sprik, M. Parrinello and R. Car, J. Chem. Phys. \textbf{99}, 9080 (1993).

\bibitem{tuckerman_jcp_1995} M. Tuckerman, K. Laasonen, M. Sprik and M. Parrinello, J. Chem. Phys. \textbf{103}, 150 (1995).

\bibitem{Ikeda_2003} T. Ikeda, M. Hirata and T. Kimura, J. Chem. Phys. {\bf 119}, 12386 (2003).

\bibitem{Heuft_2003} J. M. Heuft and E. J. Meijer, J. Chem. Phys. {\bf 119}, 11788 (2003).

\bibitem{Tongraar_2003} A. Tongraar and B. M. Rode, Phys. Chem. Chem. Phys. {\bf 5}, 357 (2003).

\bibitem{Mallik_2008} B. S. Mallik, A. Semparithi and A. Chandra, J. Chem. Phys. {\bf 129}, 194512 (2008).

\bibitem{Scipioni_2009} R. Scipioni, D. A. Schmidt and M. A. Boero, J. Chem. Phys. {\bf 130}, 024502  (2009).

\bibitem{Guardia_2009}E. Guardia, I. Skarmoutsos and M. Masia, J. Chem. Theory Comput. {\bf 5}, 1449 (2009).

\bibitem{Calemana_2011}C. Calemana, J. S. Hubb, P. J. van Maaren and D. van der Spoel, Proc. Natl. Acad. Sci. USA {\bf 108}, 6838 (2011).

\bibitem{Leung_2009} K. Leung, S. B. Rempe and O. A. von Lilienfeld, J. Chem. Phys. {\bf 130}, 204507 (2009).

\bibitem{Bankura_2013} A. Bankura, V. Carnevale and M. L. Klein, J. Chem. Phys. {\bf 138}, 014501 (2013).

\bibitem{Ge_2013}L.  Ge, L. Bernasconi and P. Hunt, Phys. Chem. Chem. Phys. {\bf 15}, 13169 (2013).

\bibitem{Zhang_2013}C. Zhang, T. A. Pham, F. Gygi and G. Galli, J. Chem. Phys. {\bf 138}, 181102 (2013).

\bibitem{Gaiduk_2014}A. P. Gaiduk, C. Zhang, F. Gygi and G. Galli, Chem. Phys. Lett. {\bf 604}, 89 (2014).

\bibitem{Car_1985}R. Car and M. Parrinello, Phys. Rev. Lett. {\bf 55}, 2471 (1985).

\bibitem{geissler_science_2001} P. L. Geissler, C. Dellago, D. Chandler, J. Hutter and M. Parrinello, Science {\bf 291}, 2121 (2001). 

\bibitem{ikeshoji_pccp_2009} T. Ikeshoji, M. Otani, I. Hamada and Y. Okamoto, Phys. Chem. Chem. Phys. \textbf{13}, 20223 (2011).

\bibitem{Zipoli_2010} F. Zipoli, R. Car, M. H. Cohen and A. Selloni, J. Am. Chem. Soc. {\bf 132}, 8593 (2010). 

\bibitem{Perdew_1992} J. P. Perdew, J. A. Chevary, S. H. Vosko, K. A. Jackson, M. R.Pederson, D. J. Singh and C. Fiolhais, Phys. Rev. B {\bf 46}, 6671 (1992). 

\bibitem{Becke_1992} A. D. Becke, J. Chem. Phys. {\bf 96}, 2155 (1992); {\bf 97}, 9173 (1992).

\bibitem{Perdew_1996} J. P. Perdew, K. Burke and M. Ernzerhof, Phys. Rev. Lett. {\bf 77}, 3865 (1996).

\bibitem{Becke_1988} A. D. Becke, Phys. Rev. A {\bf 38}, 3098 (1988).

\bibitem{Lee_1988} C. Lee, W. Yang and R. G. Parr, Phys. Rev. B {\bf 37}, 785 (1988).

\bibitem{Barone_1999} C. Adamo and V. Barone, J. Chem. Phys. {\bf 110}, 6158 (1999).

\bibitem{Todorova_2006}T. Todorova, A. P. Seitsonen, J. Hutter, I. F. Kuo and C. J. Mundy, J. Phys. Chem. B {\bf 110}, 3685 (2006).

\bibitem{DelBen_2013} M. Del Ben, M. Sch{\"{o}}nherr, J. Hutter and J. VandeVondele, J. Phys. Chem. Lett. {\bf 4}, 3753 (2013).

\bibitem{asthagiri_pre_2003} D. Asthagiri, L. R. Pratt and J. D. Kress, Phys. Rev. E \textbf{68}, 041505  (2003).

\bibitem{grossman_jcp_2004} J. C. Grossman, E. Schwegler, E.W. Draeger, F. Gygi and G. Galli, J. Chem.  Phys.  \textbf{120}, 300 (2004).

\bibitem{schwegler_jcp_2004} E. Schwegler, J. C. Grossman, F. Gygi and G. Galli, J. Chem. Phys. \textbf{121}, 5400 (2004).

\bibitem{fernandez_jcp_2004} M. V. Fern\'{a}ndez-Serra and E. Artacho, J. Chem. Phys.  \textbf{121}, 11136  (2004).

\bibitem{kuo_jpcb_2004} I. F. W. Kuo, C. J. Mundy, M. J. McGrath, J. I. Siepmann, J. VandeVondele, M. Sprik, J. Hutter, B. Chen, M. L. Klein, F. Mohamed, M. Krack and M. Parrinello, J. Phys. Chem. B \textbf{108}, 12990 (2004).

\bibitem{mcgrath_cpc_2005} M. J. McGrath, J. I. Siepmann, I. F. W. Kuo, C. J. Mundy, J. VandeVondele, J.  Hutter, F. Mohamed and M. Krack, Chem. Phys. Chem. \textbf{6}, 1894 (2005).

\bibitem{vandevondele_jcp_2005} J. VandeVondele, F. Mohamed, M. Krack, J. Hutter, M. Sprik and M. Parrinello, J. Chem. Phys. \textbf{122}, 014515 (2005).

\bibitem{sit_jcp_2005} P. H. L. Sit and N. Marzari, J. Chem. Phys. \textbf{122}, 204510 (2005).

\bibitem{fernandez_ms_2005} M. V. Fern\'{a}ndez-Serra, G. Ferlat and E. Artacho, Molecular Simulation \textbf{31}, 361 (2005).

\bibitem{mcgrath_mp_2006} M. J. McGrath, J. I. Siepmann, I. F. W. Kuo and C. J. Mundy, Mol. Phys. \textbf{104}, 3619 (2006).

\bibitem{lee_jcp_2006} H. S. Lee and M. E. Tuckerman, J. Chem. Phys. \textbf{125}, 154507 (2006).

\bibitem{lee_jcp_2007} H. S. Lee and M. E. Tuckerman, J. Chem. Phys. \textbf{126}, 164501 (2007).

\bibitem{guidon_jcp_2008} M. Guidon, F. Schiffmann, J. Hutter and J. VandeVondele, J. Chem. Phys. \textbf{128}, 214104 (2008).

\bibitem{kuhne_jctc_2009}
T. D. K\"{u}hne, M. Krack and M. Parrinello,  
J. Chem. Theory Comput.  \textbf{5}, 235 (2009).

\bibitem{mattson_jctc_2009}
A. E. Mattsson and T. R. Mattsson,  
J. Chem. Theory Comput.  \textbf{5}, 887  (2009).

\bibitem{yoo_jcp_2009}
S. Yoo, X. C. Zeng and S. S. Xantheas,  
J. Chem. Phys.  \textbf{130}, 221102  (2009).

\bibitem{bartok_prb_2013}
A. P. Bart\'{o}k, M. J. Gillan, F. R. Manby and G. Cs\'{a}nyi,  
Phys. Rev. B  \textbf{88}, 054104 (2013).

\bibitem{alfe_jcp_2013}
D. Alf\`e, A. P. Bart\'{o}k, G. Cs\'{a}nyi and M. J. Gillan,  
J. Chem. Phys.  \textbf{138}, 221102 (2013).

\bibitem{lin_jpcb_2009}
I. C. Lin, A. P. Seitsonen, M. D. Coutinho-Neto, I. Tavernelli and U.  Rothlisberger,  
J. Phys. Chem. B  \textbf{113}, 1127 (2009).

\bibitem{jonchiere_jcp_2011}
R. Jonchiere, A. P. Seitsonen, G. Ferlat, A. M. Saitta and R. Vuilleumier,  
J.  Chem. Phys.  \textbf{135}, 154503 (2011).

\bibitem{wang_jcp_2011}
J. Wang, G. Rom\'{a}n-P\'{e}rez, J. M. Soler, E. Artacho and M. V. Fern\'{a}ndez-Serra,  
J. Chem. Phys.  \textbf{134}, 024516 (2011).

\bibitem{zhang_jctc_2011}
C. Zhang, J. Wu, G. Galli and F. Gygi,  
J. Chem. Theory Comput.  \textbf{7}, 3054 (2011).

\bibitem{mogelhoj_jpcb_2011}
A. M{\o}gelh{\o}j, A. K. Kelkkanen, K. T. Wikfeldt, J. Schi{\o}tz, J. J.
  Mortensen, L. G. M. Pettersson, B. I. Lundqvist, K. W. Jacobsen, A. Nilsson and
  J. K. N{\o}rskov,  J. Phys. Chem. B  \textbf{115}, 14149 (2011).

\bibitem{lin_jctc_2012}
I. C. Lin, A. P. Seitsonen, I. Tavernelli and U. Rothlisberger,  
J. Chem. Theory  Comput.  \textbf{8}, 3902 (2012).

\bibitem{yoo_jcp_2012}
S. Yoo and S. S. Xantheas,  J. Chem. Phys.  \textbf{134}, 121105 (2011).

\bibitem{schmidt_jpcb_2009}
J. Schmidt, J. VandeVondele, I. F. W. Kuo, D. Sebastiani, J. I. Siepmann, J.
  Hutter and C. J. Mundy,  J. Phys. Chem. B  \textbf{113}, 11959 (2009).

\bibitem{ma_jcp_2012}
Z. Ma, Y. Zhang and M. E. Tuckerman,  J. Chem. Phys.  \textbf{137}, 044506 (2012).

\bibitem{Zhang_2011}C. Zhang, D. Donadio, F. Gygi and G. Galli,
J. Chem. Theory Comput. {\bf 7}, 1443 (2011).

\bibitem{feibelman_pccp_2008}
P. J. Feibelman,  Phys. Chem. Chem. Phys.  \textbf{10}, 4688 (2008).

\bibitem{santra_prl_2011}
B. Santra, J. Klime\v{s}, D. Alf\`e, A. Tkatchenko, B. Slater, A. Michaelides,
  R. Car and M. Scheffler,  Phys. Rev. Lett.  \textbf{107}, 185701 (2011).

\bibitem{santra_jcp_2013}
B. Santra, J. Klime\v{s}, A. Tkatchenko, D. Alf\`e, B. Slater, A. Michaelides,
  R. Car and M. Scheffler,  J. Chem. Phys.  \textbf{139}, 154702 (2013).

\bibitem{labat_jcc_2011}
F. Labat, C. Pouchan, C. Adamo and G.E. Scuseria,  J. Comput. Chem.
  \textbf{32}, 2177 (2011).
  
\bibitem{kambara_pccp_2012}
O. Kambara, K. Takahashi, M. Hayashi and J.L. Kuo,  Phys. Chem. Chem. Phys.
  \textbf{14}, 11484 (2012).

\bibitem{murray_prl_2012}
E. D. Murray and G. Galli,  Phys. Rev. Lett.  \textbf{108}, 105502 (2012).

\bibitem{fang_prb_2013}
Y. Fang, B. Xiao, J. Tao, J. Sun and J. P. Perdew,  Phys. Rev. B  \textbf{87},
  214101 (2013).

\bibitem{macher_jcp_2014}
M. Macher, J. Klime\v{s}, C. Franchini and G. Kresse,  J. Chem. Phys.
  \textbf{140}, 084502 (2014).

\bibitem{Shore_1977}H. B. Shore, J. H. Rose and E. Zaremba 
Phys. Rev. B {\bf 15}, 2858 (1977).

\bibitem{perdew_prb_1981}
J. Perdew and A. Zunger,  Phys. Rev. B  \textbf{23}, 5048 (1981).

\bibitem{chen_prl_2003}
B. Chen, I. Ivanov, M. L. Klein and M. Parrinello,  Phys. Rev. Lett.
  \textbf{91}, 215503 (2003).
  
\bibitem{ceriotti_pnas_2013}
M. Ceriotti, J. Cuny, M. Parrinello and D. E. Manolopoulos,  Proc. Natl. Acad.
  Sci. USA  \textbf{110}, 15591 (2013).

\bibitem{morrone_prl_2008}
J.A. Morrone and R. Car,  Phys. Rev. Lett.  \textbf{101}, 017801 (2008).

\bibitem{soper_prl_2008}
A. Soper and C. J. Benmore,  Phys. Rev. Lett.  \textbf{101}, 065502 (2008).

\bibitem{ceriotti_prl_2012}
M. Ceriotti and D. E. Manolopoulos,  Phys. Rev. Lett.  \textbf{109}, 100604
  (2012).

\bibitem{paesani_jcp_2007}
F. Paesani, S. Iuchi and G. A. Voth,  J. Chem. Phys.  \textbf{127}, 074506
  (2007).

\bibitem{fanourgakis_jcp_2006}
G. S. Fanourgakis, G. K. Schenter and S. S. Xantheas,  J. Chem. Phys.
  \textbf{125}, 141102 (2006).

\bibitem{santra_jcp_2007}
B. Santra, A. Michaelides and M. Scheffler,  J. Chem. Phys.  \textbf{127},
  184104 (2007).

\bibitem{santra_jcp_2008}
B. Santra, A. Michaelides, M. Fuchs, A. Tkatchenko, C. Filippi and M.
  Scheffler,  J. Chem. Phys.  \textbf{129}, 194111 (2008).
  
  \bibitem{santra_jcp_2009}
B. Santra, A. Michaelides and M. Scheffler,  J. Chem. Phys.  \textbf{131},
  124509 (2009).

\bibitem{wang_jcp_2010}
F. F. Wang, G. Jenness, {W. A. Al-Saidi} and K. D. Jordan,  J. Chem. Phys.
  \textbf{132}, 134303 (2010).

\bibitem{gillan_jcp_2012}
M. J. Gillan, F. R. Manby, M. D. Towler and D. Alf\`e,  J. Chem. Phys.
  \textbf{136}, 244105 (2012).

\bibitem{erba_jpcb_2009} A. Erba, S. Casassa, L. Maschio and C. Pisani,  J. Phys. Chem. B \textbf{113}, 2347 (2009).

\bibitem{Distasio2014} R. A. DiStasio Jr., B. Santra, Z. Li, X. Wu and R. Car, J. Chem. Phys. \textbf{141}, 084502 (2014).

\bibitem{perdew_jcp_1996} J. P. Perdew, M. Ernzerhof and K. Burke, J. Chem. Phys. \textbf{105}, 9982 (1996).

\bibitem{Tkatchenko_2009} A. Tkatchenko and M. Scheffler, Phys. Rev. Lett. {\bf 102}, 073005 (2009).

\bibitem{Skinner2013} L. B. Skinner, C. Huang, D. Schlesinger, L. G. M. Pettersson, A. Nilsson and C. J. Benmore, J. Chem. Phys. {\bf 138}, 074506 (2013).

\bibitem{distasio_unpublished} R. A. DiStasio Jr., B. Santra, {H.-Y. Ko} and R. Car, \textit{to be published}.

\bibitem{landau_sm_book_1969} L. D. Landau and E. M. Lifshitz, \emph{Statistical Physics}, \emph{Course of Theoretical Physics}, Vol.~5 (Pergamon Press, Oxford, 1969).
  
\bibitem{QE-2009} P. Giannozzi, S. Baroni, N. Bonini, M. Calandra, R. Car, C. Cavazzoni, D. Ceresoli, G. L. Chiarotti, M. Cococcioni, I. Dabo, {A. Dal Corso}, {S. de Gironcoli}, S. Fabris, G. Fratesi, R. Gebauer, U. Gerstmann, C. Gougoussis, A. Kokalj, M. Lazzeri, {L. Martin-Samos}, N. Marzari, F. Mauri, R. Mazzarello, S. Paolini, A. Pasquarello, L. Paulatto, C. Sbraccia, S. Scandolo, G. Sclauzero, A. P. Seitsonen, A. Smogunov, P. Umari and R. M. Wentzcovitch, J. Phys.: Condens. Matter \textbf{21}, 395502 (2009).

\bibitem{marzari_prb_1997} N. Marzari and D. Vanderbilt,  Phys. Rev. B \textbf{56}, 12847 (1997).

\bibitem{wu_prb_2009} X. Wu, A. Selloni and R. Car,  Phys. Rev. B \textbf{79}, 085102 (2009).

\bibitem{ko_unpublished} H.-Y. Ko, B. Santra, R. A. DiStasio Jr., L. Kong, Z. Li, X. Wu, and R. Car, ``Enabling Hybrid Density Functional Theory Calculations on Large-Scale Condensed-Phase Materials,'' \textit{to be published}.


\bibitem{troullier_prb_1991} N. Troullier, J. L. Martins, Phys. Rev. B {\bf 43} 1993 (1991).

\bibitem{Nose_1984} S. A. Nos\'e, J. Chem. Phys. {\bf 81}, 511 (1984).

\bibitem{Hoover_1985}W. G. Hoover, Phys. Rev. A {\bf 31}, 1695 (1985).

\bibitem{martyna_jcp_1992} G. J. Martyna, M. L. Klein and M. Tuckerman,  J. Chem. Phys.  \textbf{97}, 2635 (1992).

\bibitem{Becke_1993}A. D. Becke, J. Chem. Phys. {\bf 98}, 1372 (1993).

\bibitem{cp2k} The CP2K developers group. http://cp2k.org/ ( 2000-2014) .

\bibitem{Goedecker_1996} S. Goedecker, M. Teter and J. Hutter, Phys. Rev. B {\bf 54}, 1703 (1996).

\bibitem{Yang_Science} A. J. Cohen, P. Mori-S\,anchez, and W. Yang, Science {\bf 321}, 792 (2008).

\bibitem{Adriaanse_2012} C. Adriaanse, J. Cheng, V. Chau, M. Sulpizi, J. VandeVondele and M. Sprik, J. Phys. Chem. Lett. {\bf 3}, 3411 (2012).


\bibitem{Liu_2015} X. Liu, J. Cheng and M. Sprik, J. Phys. Chem. B {\bf 119}, 1152 (2015).

\bibitem{Cheng_2014} J. Cheng, X. Liu, J. VandeVondele, M. Sulpizi and M. Sprik, Acc. Chem. Res. {\bf 47}, 3522 (2014).

\bibitem{Pham_2014} T. A. Pham, C. Zhang, E. Schwegler and G. Galli, Phys. Rev. B {\bf 89}, 060202(R) (2014).

\bibitem{Viktor_2013} V. Atalla, M. Yoon, F. Caruso, P. Rinke, and M. Scheffler, Phys. Rev. B {\bf 88}, 165122 (2013).


\bibitem{Swartz_2013} C. Swartz and X. Wu, Phys. Rev. Lett. {\bf 111}, 087801 (2013).

\bibitem{Private} Private communication with Prof. Roberto Car, Princeton University (2015).

\bibitem{Skone2014} J. H. Skone, M. Govoni, and G. Galli, Phys. Rev. B {\bf 89}, 195112 (2014).

\bibitem{tkatchenko_prl_2012} A. Tkatchenko, R. A. DiStasio Jr., R. Car, and M. Scheffler, Phys. Rev. Lett. {\bf 108}, 236402 (2012).

\bibitem{distasio_pnas_2012} R. A. DiStasio Jr., O. A. von Lilienfeld, and A. Tkatchenko, Proc. Natl. Acad. Sci. USA {\bf 109}, 14791 (2012).

\bibitem{ambrosetti_jcp_2014} A. Ambrosetti, A. M. Reilly, R. A. DiStasio Jr., and A. Tkatchenko, J. Chem. Phys. {\bf 140}, 18A508 (2014).

\bibitem{distasio_jpcm_2014} R. A. DiStasio Jr., V. V. Gobre, and A. Tkatchenko, J. Phys.: Condens. Matter {\bf 26}, 213202 (2014).

\bibitem{GW1} L. Hedin, Phys. Rev. \textbf{139}, A796 (1965).

\bibitem{RPA1} D. Bohm and D. Pines. Phys. Rev. \textbf{82}, 625 (1951).

\bibitem{RPA2} D. Pines and D. Bohm, Phys. Rev. \textbf{85}, 338 (1952).

\bibitem{RPA3} D. Bohm and D. Pines, Phys. Rev. \textbf{92}, 609 (1953).


\end{thebibliography}
\end{document}